\documentclass[aps,prl,twocolumn,superscriptaddress,showpacs]{revtex4}

\usepackage{graphicx}

\begin{document}

\title{Colossal magnetoresistance without phase separation: Disorder-induced spin glass state and nanometer scale orbital-charge correlation in half doped manganites}

\author{R. Mathieu\cite{cryo}}

\affiliation{Spin Superstructure Project (ERATO-SSS), JST, AIST Central 4, Tsukuba 305-8562, Japan}

\author{D. Akahoshi}
\affiliation{Correlated Electron Research Center (CERC), AIST Central 4, Tsukuba 305-8562, Japan}

\author{A. Asamitsu}
\affiliation{Spin Superstructure Project (ERATO-SSS), JST, AIST Central 4, Tsukuba 305-8562, Japan}
\affiliation{Cryogenic Research Center (CRC), University of Tokyo, Bunkyo-ku, Tokyo 113-0032, Japan}

\author{Y. Tomioka}
\affiliation{Correlated Electron Research Center (CERC), AIST Central 4, Tsukuba 305-8562, Japan}

\author{Y. Tokura}
\affiliation{Spin Superstructure Project (ERATO-SSS), JST, AIST  Central 4, Tsukuba 305-8562, Japan}
\affiliation{Correlated Electron Research Center (CERC), AIST Central 4, Tsukuba 305-8562, Japan}
\affiliation{Department of Applied Physics, University of Tokyo, Tokyo 113-8656, Japan}

\date{\today}

\begin{abstract}
The magnetic and electrical properties of high quality single crystals of $A$-site disordered (solid solution) Ln$_{0.5}$Ba$_{0.5}$MnO$_3$ are investigated near the phase boundary between the spin glass insulator and colossal-magnetoresistive ferromagnetic metal, locating near Ln = Sm. The temperature dependence of the ac-susceptibility and the x-ray diffuse scattering of Eu$_{0.5}$Ba$_{0.5}$MnO$_3$ are analyzed in detail. The uniformity of the random potential perturbation in Ln$_{0.5}$Ba$_{0.5}$MnO$_3$ crystals with small bandwidth yields, rather than the phase separation, an homogeneous short ranged charge/orbital order which gives rise to a nearly-atomic spin glass state. Remarkably, this microscopically disordered ``CE-glass'' state alone is able to bring forth the colossal magnetoresistance.
\end{abstract}


\maketitle
The phase diagram of colossal magnetoresistive (CMR) perovskite manganites, as well as many strongly correlated electron systems, is multicritical, involving competing spin, charge/orbital, and lattice orders\cite{competition,open}. For example in the half-doped Ln$_{0.5}$Ba$_{0.5}$MnO$_3$  (Ln being a rare earth cation), the charge/orbital ordered (CO/OO) insulating (favored by small Ln cations) and ferromagnetic (FM) metallic (larger Ln cations) states compete with each other, and bicritically meet near Ln = Nd\cite{Akahoshi}. The meeting point is actually multicritical, as the $A$-type antiferromagnetic (AFM) instability also exists\cite{Akahoshi}. In the presence of quenched disorder, namely when the perovskite $A$-sites are solid solution of Ln and Ba (termed ``$A$-site disorder''), the phase diagram becomes asymmetric. The FM phase transition is still observed near the critical point, even though the Curie temperature $T_c$ is steeply diminished. The long-range CO/OO state is, on the other hand, completely suppressed and only short-range CO/OO correlation is observed. This phase corresponds in the spin sector to a spin glass (SG) state, which, as we will show in this article, is not related to some macroscopic phase separation\cite{Raveau,Kimura}, but results from the frustration and magnetic disorder microscopically introduced within this ``CE-glass''\cite{Dagotto} state. The degree of quenched disorder can be controlled by modifying the mismatch of the constituent $A$-site cations\cite{Attfield}. Recent pressure experiments show that the asymmetric phase diagram of disordered Ln$_{0.5}$Ba$_{0.5}$MnO$_3$ (LnBMO) is solely determined by the bandwidth $W$ variation, for a fixed degree of disorder\cite{Takeshita}. It thus implies that quenched disorder affects the CO/OO phase more aggressively than the FM state. These experimental findings could be accounted for by the theoretical model considering multicritical fluctuations between the FM and CO/OO phases\cite{Nagaosa}, in presence of disorder. The calculations also predict an enhanced electronic localization above $T_c$ near the critical region, suggesting that large CMR effects may emerge\cite{Nagaosa}.

We thus investigate the magnetic and electrical properties of high-quality single crystals of $A$-site disordered Ln$_{0.5}$Ba$_{0.5}$MnO$_3$, around the SG insulator/FM metal phase boundary (near  Ln = Sm). The temperature $T$ and frequency $f$ dependence of the ac-susceptibility of Eu$_{0.5}$Ba$_{0.5}$MnO$_3$ (Ln = Eu, EBMO) is analyzed in detail. The low-temperature EBMO resembles canonical spin glasses, and characteristic phenomena such as aging, memory, and rejuvenation are observed at low temperatures. This spin glass state, and thus the short-range orbital-order revealed by the x-ray diffuse scattering, appear homogeneous down to the nanometer scale. This suggests that, in Ln$_{0.5}$Ba$_{0.5}$MnO$_3$ crystals with relatively small bandwidth, the effect of the disorder is microscopic, and does not induce the phase separation. The observed ``CE-glass'' state appears to contain all the building blocks for the colossal magnetoresistance phenomenon.
 
Single crystals of $A$-site disordered Ln$_{0.5}$Ba$_{0.5}$MnO$_3$  (Ln = Eu, Sm, Nd, Pr, and La) were grown by the floating zone method\cite{Akahoshi}. The phase-purity of the crystals and the perfect disorder (solid solution) of the Ln/Ba cations on the $A$-site was checked by x-ray diffraction\cite{Akahoshi}. Single crystal x-ray diffraction was performed at selected temperatures from 300 K down to 30 K using an imaging plate system equipped with a closed cycle helium refrigerator. The magnetization and ac-susceptibility $\chi$($T,\omega=2\pi f$) data were recorded on a MPMSXL SQUID magnetometer equipped with the ultra low-field option (low frequencies) and a PPMS6000 (higher frequencies), after carefully zeroing or compensating the background magnetic fields of the systems. Additional phase corrections were performed for some frequencies. The transport properties were measured using a standard four-probe method.

\begin{figure}[h]
\includegraphics[width=0.46\textwidth]{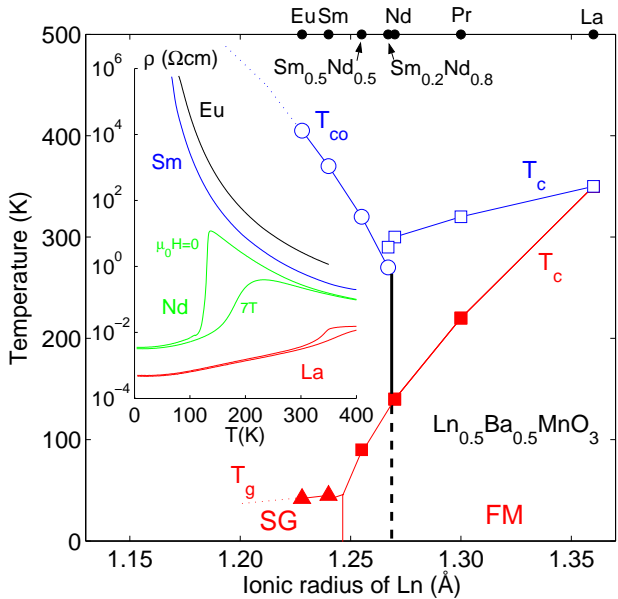}
\caption{(color online) Electronic phase diagram for the $A$-site ordered and disordered Ln$_{0.5}$Ba$_{0.5}$MnO$_3$ as a function of the ionic radius of Ln. Ln = Eu, Sm, Sm$_{1-y}$Nd$_y$ ($y$ = 0.5, 0.8), Nd, Pr, and La, reproduced from Ref. {\protect \onlinecite{Akahoshi}}. For the Ln/Ba ordered compounds, the AFM transitions below $T_c$ (A-type AFM; Ln = Nd and Pr) and T$_{co}$ (accompanying a rearrangement of the CO/OO state; Ln = Eu, Sm, and Sm$_{0.2}$Nd$_{0.8}$), are not shown for clarity. No AFM transition is observed for the $A$-site disordered compounds. $T_c$ is the Curie temperature,  $T_{co}$ the charge-ordering temperature, $T_g$ the spin glass (SG) phase transition obtained from dynamical scaling. The inset shows the temperature dependence of the resistivity of the sample with  Ln = Eu, Sm, Nd, and La, measured for $\mu_0H$ = 0 and 7 T.}
\label{fig-diag}
\end{figure}

Figure~\ref{fig-diag} shows the asymmetric phase diagram of the $A$-site disordered LnBMO using filled symbols. Results for the  $A$-site ordered samples\cite{Akahoshi} are shown using open symbols for comparison.  For large cations, the FM phase remains, albeit with reduced $T_c$. The $T_c$ is steeply suppressed, as the Ln ion size decreases. For smaller Ln cations, a spin glass phase with non-equilibrium dynamics emerges. The inset of Fig.~\ref{fig-diag} shows the resistivity of the samples with Ln = Eu, Sm, Nd, and La measured in zero and applied magnetic field. Eu$_{0.5}$Ba$_{0.5}$MnO$_3$ and Sm$_{0.5}$Ba$_{0.5}$MnO$_3$ (SBMO) show insulator-like resistivity in 0 and 7 T; larger magnetic fields may be required to induce metallicity and magnetoresistance.

\begin{figure}[h]
\includegraphics[width=0.46\textwidth]{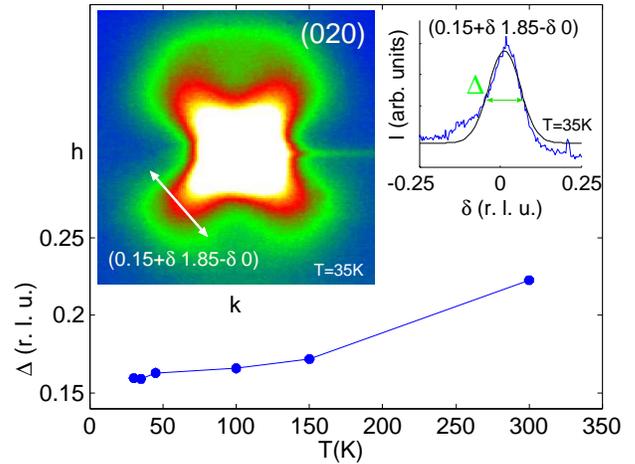}
\caption{(color online) 
Temperature dependence of the width of the x-ray diffuse scattering $\Delta$ around the (0 2 0) Bragg peak (in the cubic setting; shown in the left inset) in EBMO. $\Delta$ is estimated as the full width at half maximum of the Gaussian fit of [($h$=0.15+$\delta$ $k$=1.85-$\delta$ 0), -0.25$\leq \delta \leq$0.25] diffuse scattering profiles, as shown in the right inset.}
\label{fig-xr}
\end{figure}
\begin{figure}[h]
\includegraphics[width=0.46\textwidth]{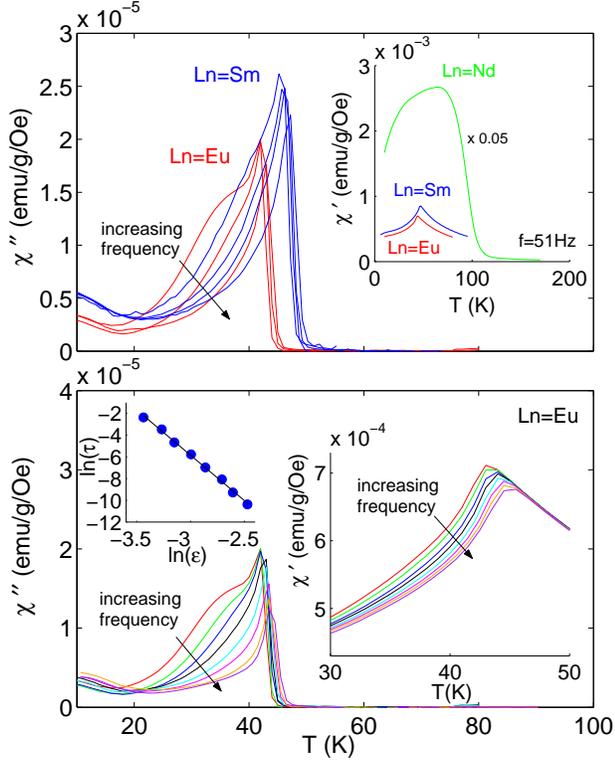}
\caption{(color online) Upper panel: Temperature ($T$) dependence of the out-of-phase component of the ac-susceptibility  $\chi''$($T,\omega$) for the compounds with Ln = Eu and Sm; $h$ (field amplitude) = 4 Oe, $f$ ($=\omega / 2\pi$, frequency)  = 1.7, 17, and 170 Hz for Ln = Eu, $h$ = 10 Oe, $f$ =100, 500, 1000, and 5000 Hz for Ln = Sm. The $T$-dependence of the in-phase component  $\chi'$($T,\omega$) for Ln = Eu, Sm, and Nd is shown in inset. The frequency dependence of $\chi$($T,\omega$) of EBMO is studied in more detail in the lower panel: the $T$-dependence of $\chi''$($T,\omega$) is shown in main frame, while $\chi'$($T,\omega$) is shown in inset; $h$ = 4 Oe, $f$ = 1.7, 5.1, 17, 51, 170, 510, 1700, and 5100 Hz. The left inset shows the dynamical scaling of $\tau$($T_f$) = $t_{obs}$ with the reduced temperature $\epsilon$ for  $T_g$ = 42 K, $z\nu$ = 7.6 and $\tau_0$ $\sim$ 1.7$\times$10$^{-13}$ s. }
\label{fig-ac}
\end{figure}
However, a FM metallic state can be induced by application of hydrostatic pressure\cite{Takeshita}. LnBMO samples with larger Ln cations occupy a ``critical zone'' near the multicritical point of the phase diagram. For example, Nd$_{0.5}$Ba$_{0.5}$MnO$_3$ (NBMO) shows a metallic ground state, as well as a large CMR effect in the vicinity of $T_c$. It is known that the long-range CO/OO order can be locally hindered by impurities substituting the Mn-sites, as in Cr doped Pr$_{0.5}$Ca$_{0.5}$MnO$_3$\cite{Raveau} or Nd$_{0.5}$Ca$_{0.5}$MnO$_3$\cite{Kimura}. For Cr concentrations $\leq$ 5 \%, the CO/OO coherence remains on relatively large length scales, yielding separation of FM and CO/OO phases\cite{Kimura}, and associated percolative metal-to-insulator transition upon field application. In the present case, on the contrary, the Ln/Ba solid solution on the perovskite $A$-sites induces a global randomness in the potential, which breaks the CO/OO coherence down to the nanometer scale. The short-range nature of the charge-orbital correlation is evidenced by the x-ray diffuse scattering\cite{diffuse} observed at all temperatures in EBMO, as illustrated in Fig.~\ref{fig-xr}. The intensity of the diffuse scattering around the (0 2 0) Bragg peak increases with decreasing temperature, as observed in the similar Ln$_{0.55}$Sr$_{0.45}$MnO$_3$ system\cite{Tomioka-RSMO-PCSMO}. The width ($\Delta$) of the diffuse scattering profile has a relatively weak temperature dependence, and saturates at low temperatures around  $\Delta \sim$ 0.16 r.l.u.. Considering a pseudocubic lattice parameter of 0.38 nm, this corresponds to a charge-orbital coherence length $\xi$ $\sim$ 2 nm. Thus, as in Ln$_{0.55}$Sr$_{0.45}$MnO$_3$\cite{Tomioka-RSMO-PCSMO}, no macroscopic phase separation is discerned, and the observed CMR effect of Nd$_{0.5}$Ba$_{0.5}$MnO$_3$ may reflect the augmented response to magnetic fields of systems with enhanced CO/OO fluctuations\cite{Nagaosa}.  Away from the SG/FM phase boundary, Pr$_{0.5}$Ba$_{0.5}$MnO$_3$ (not shown) and La$_{0.5}$Ba$_{0.5}$MnO$_3$  (LBMO) exhibit the conventional metallicity and magnetoresistance around $T_c$, as seen in the inset of Fig.~\ref{fig-diag}.

The top panel of Fig.~\ref{fig-ac} shows the temperature dependence of the ac-susceptibility of the EBMO, SMBO and NBMO sample. As seen in the inset, the in-phase component $\chi'$($T,\omega$) of the susceptibility of NBMO is typical of a FM material with a sharp ferromagnetic transition. The SBMO and EBMO instead show reduced magnetization and broad transitions, similar to those of disordered systems. The out-of-phase component $\chi''$($T,\omega$) of the susceptibility shows no frequency dependence at the onset of FM in the case of NBMO. In SBMO and EBMO, as shown in the main frame of Fig.~\ref{fig-ac}, $\chi''$($T,\omega$) exhibits a fairly large $f$-dependence. $\chi$($T,\omega$) is recorded in a small ac magnetic field, allowing us to probe the system (and its dynamics) without affecting it; the linear response of the system was confirmed.
\begin{figure}[h]
\includegraphics[width=0.46\textwidth]{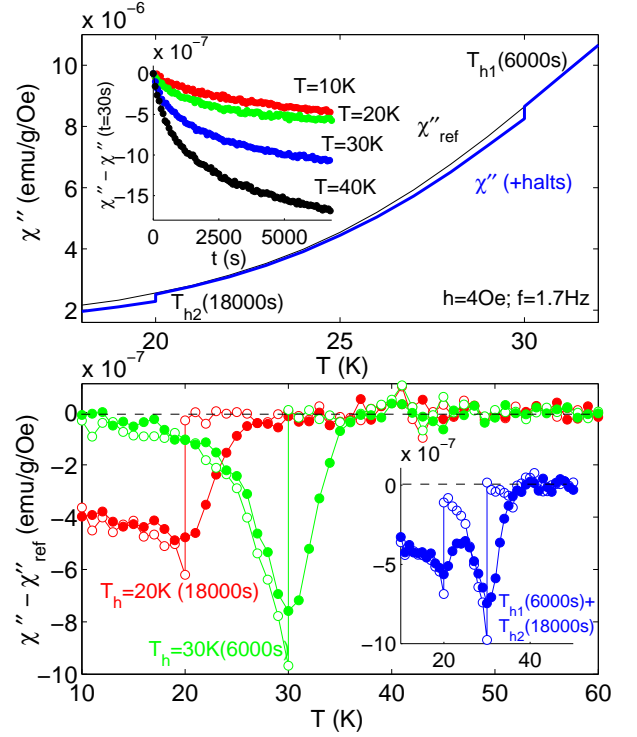}
\caption{(color online) Upper panel: Out-of-phase component of the ac-susceptibility vs temperature, measured on continuous cooling (thin line), and including two halts at $T_{h1}$ = 30 K and  $T_{h2}$ = 20 K (thicker line). The inset shows the relaxation of $\chi''$($T,\omega$) at constant temperatures below the glass transition of EBMO. $f$ = 1.7 Hz, $h$ = 4 Oe. Lower panel: Difference plots of $\chi''$($T,\omega$) curves measured on cooling (open symbols) and subsequent re-heating (closed symbols) including halts, subtracted with their respective references measured on continuous cooling and heating. Results of single halts at $T_{h}$ = 30 K and  $T_{h}$ = 20 K are shown in the main frame, while a double memory experiment with halts at both $T_{h1}$ = 30 K and  $T_{h2}$ = 20 K is shown in inset for comparison.}
\label{fig-memo}
\end{figure}
We analyze in detail the $T$- and $f$-dependence of $\chi''$($T,\omega$) of EBMO shown in the lower panel of Fig.~\ref{fig-ac}. Eu$^{3+}$ is smaller than Sm$^{3+}$, so that EBMO is conveniently away from the phase boundary, and Eu has no large $4f$ moment affecting $\chi$($T,\omega$). The peak observed in $\chi''$($T,\omega$) at all $f$ is relatively sharp, which suggests that the system homogeneously undergoes a phase transition. Each frequency corresponds to an observation time $t_{obs} = 1/\omega$ characteristic of the measurement. One can define from each susceptibility curve a frequency dependent freezing temperature $T_f$($\omega$), below which the longest relaxation time of the system exceeds $t_{obs}$, and the system is out-of-equilibrium. The inset in the lower panel of Fig.~\ref{fig-ac} shows the scaling of $\tau$($T_f$) = $t_{obs}$ with the reduced temperature $\epsilon=(T_f(f)-T_g)/T_g$ ($T_g$ is the spin glass phase transition temperature) using a conventional critical slowing-down power-law relation\cite{dynscal}, linear in log-log axes.  A good scaling is obtained for $T_g$ = 42 $\pm$ 1 K, $z\nu$ = 8 $\pm 1$ and $\tau_0$ $\sim$ 10$^{-13 \pm 1}$ s, which indicates that the time necessary to reach equilibrium becomes longer and longer when approaching $T_g$ = 42 K, and the relaxation time diverges at $T_g$ as $\tau/\tau_0$ = $\epsilon^{-z\nu}$. $z$ and $\nu$ are critical exponents, and  $\tau_0$ represents the microscopic flipping time of the fluctuating entities, which in the present case is close to that of the microscopic spin flip time (10$^{-13}$ s). The value of the product $z\nu$ is similar to those of ordinary atomic SG\cite{exponents}. This indicates that LnBMO crystals with small Ln cations undergo a true spin glass phase transition\cite{note-sm}, and that the low temperature SG phase is homogeneously disordered, down to the nanometer scale. This can be expected considering that without quenched disorder, the long range CO/OO state consists of ferromagnetic zigzag chains running along the [110] direction (cubic setting) which are coupled antiferromagnetically (so-called CE-type)\cite{Jirak}. Thus, in this uniformely disordered case, the fragmentation of the zigzag chains down to the nanometer scale (as revealed by the x-ray diffuse scattering) causes the mixture of AFM and FM bonds on these near atomic length scales.

EBMO exhibits dynamical features typical of SG, such as aging, memory, and rejuvenation. These phenomena can be observed employing specific cooling protocols while recording $\chi$($T,\omega$), and explained using a convenient real space picture known as the droplet model\cite{droplet,ghost}. In the droplet model, the slow dynamics is related to the slow rearrangement of domain walls of the SG phase by thermal activation. After a quench from the paramagnetic phase into the spin-glass phase the system is trapped in a random non-equilibrium spin configuration which slowly equilibrates or ages. This is illustrated in the top panel of Fig.~\ref{fig-memo}. As seen in inset, $\chi''$($T,\omega$) decreases with time at constant temperature, after a quench from a reference temperature above $T_g$. This reflects the aging process, in which the number of droplets of relaxation time $1/\omega$ (probed by the ac-excitation of frequency $\omega$) decays as equilibrium domains are growing. The aging is also observed if, as shown in the main frame of Fig.~\ref{fig-memo}, $\chi''$($T,\omega$) is recorded against temperature performing halts during the cooling. $\chi''$($T,\omega$) is recorded vs temperature on cooling and heating, either continuously changing the temperature (reference curves), as well as making a halt for $t_h$ = 6000 s at $T_h$ = 30 K or for $t_h$ = 18000 s at $T_h$ = 20 K (single memory experiments), or for both $t_{h1}$ = 6000 s at $T_{h1}$ = 30 K, and  $t_{h2}$ = 18000 s at $T_{h2}$ = 20K (double memory experiments). The cooling (heating) curves measured while (after) performing the halts are subtracted from their corresponding reference curves, and plotted in the bottom panel of Fig.~\ref{fig-memo}. $\chi''$($T,\omega$) relaxes downward at $T_{h1}$ and $T_{h2}$. These agings are recovered on re-heating, as the spin configuration established during the equilibration is frozen-in upon cooling below $T_h$, and only affected on short length-scales during the re-heating\cite{ghost}. In other words, the systems keeps memory of the larger domains equilibrated during the halt. However, the equilibrium configurations at distinct temperatures are similar only on short length scales, inside a so-called overlap\cite{ghost} region. The memory of the equilibration at $T_h$ is thus observed only in a finite temperature range around $T_h$, defining ``memory dips'' with a finite width in the difference plots shown in the bottom panel of Fig.~\ref{fig-memo}. Outside this temperature range, $\chi''$($T,\omega$) recovers its reference level and the system is rejuvenated. Paradoxically\cite{ghost}, the domain state equilibrated during the halt at $T_h = T_{h1}$ survives the spin reconfiguration occurring during a second halt at $T_{h2} < T_{h1}$ on shorter length scales, an two memory dips are observed at $T_{h1}$ and $T_{h2}$, as seen in the lower panel of Fig.~\ref{fig-memo}.

In summary, we have studied $A$-site disordered Ln$_{0.5}$Ba$_{0.5}$MnO$_3$ single crystals near the spin glass insulator/ferromagnetic metal phase boundary, located near Ln = Sm. Analyses of X-ray diffuse scattering and ac-susceptibility measurements reveal that the crystals with small bandwidth behave like canonical atomic spin glasses, and suggest a short-range orbital order and glassy spin state homogeneous down to the nanometer scale. This indicates that the phase separation occurs only in crystal with local defects or impurities (such as $B$-site dopants), or equivalently, that no phase separation on a micron-scale occurs in high-quality single crystals, when Ln and Ba are ordered or otherwise perfectly disordered, by no or uniform random potential. In the latter case, the spin and orbital configurations constitute a perfect matrix for the CMR effect to originate in, as illustrated by the gigantic response to magnetic fields of the narrow bandwidth crystals.


\end{document}